\documentclass[aps,prl,twocolumn,showpacs,preprintnumbers,superscriptaddress]{revtex4}
\usepackage{amsmath}
\usepackage{amssymb}
\usepackage{graphicx}
\usepackage{dcolumn}
\usepackage{bm}

\input{tcilatex}

\begin{document}

\title{More Detailed Descriptions of Locality and Realism}
\author{Ping-Xing Chen}
\affiliation{Department of Physics, Hong Kong Baptist University, Kowloon Tong, Hong
Kong, China}
\affiliation{Department of Physics, Science College, National University of Defense
Technology, Changsha, 410073, People's Republic of China}
\author{Shi-Yao Zhu}
\affiliation{Department of Physics, Hong Kong Baptist University, Kowloon Tong, Hong
Kong, China}
\date{\today}

\begin{abstract}
Many experiments have shown that locality-realism theory is at variance with
quantum mechanics predictions. Although locality and realism, which are two
different conceptions, are given respective definition, the descriptions of
the both are a little of abstract when they are applied to real experimental
situations. The abstract descriptions result in difficulty for one to judge
whether the variance come from locality or realism or both. Here we provide
more detailed descriptions of locality and realism, and show that any system
being in a pure state or a non-maximally mixed state has property of
non-realism. We also present experimental schemes feasible under current
technologies to test the non-locality realsim. The connections between
non-locality and entanglement and correlation are also discussed.
\end{abstract}

\maketitle

The local realism theory (LRT) has been proven to be true in classical word.
In quantum word, however, LRT is at variance with quantum mechanics
predicitions. This variance was pointed out first by Einstein, Podolsky and
Rosen (EPR) \cite{epr}. But EPR think LRT should be accepted, and the
theorey of quantum mechanics (QM) is incomplete.They believe that a complete
theory of QM is possible. later, a possible complete theory of QM,
hidden-variable model (HVM), was put forward. In 1964, Bell shown that the
value of a certain combination of correlation on two distant systems cannot
be higher than a value if we accept LRT and the HVM \cite{bell}. Bell's
theorem provided a possibility for one to judge experimentally whether QM is
a LRT added by the HVM or not. Later, some improved versions of Bell's
theorem have been put forward, such as CHSH inequality and so on \cite%
{chsh,son,mer,cab}. Greenberger, Horme and Zeilinger (GHZ) \cite{ghz}
present a theorem without inequality which showed that a certain correlation
of the quantum systems may have conflict with LRT. There are many different
versions of GHZ theorem, such as some versions in Refs.\cite{hard,chen,cab1}.

LRT has two assumptions \cite{epr, zeil}: realism and locality. Realism
claims that all measurement outcomes depend on pre-existing properties of
the system that are independent of the measurement while locality claims
that there are no influences between events in space-like separated systems.
Up to now, all experiments motivated by these Bell' theorems are completely
consistent with QM's predictions \cite{bou, pan,pan1,zhao}, and so deny the
LRT in quantum word. The failure of LRT means at least that one of two
assumptions fails. Bell' theorem and its improved versions only showed that
LRT is not consistent with the QM's results, but they do not tell us that
both locality or realism or both result in this inconsistence. Does there
exists non-locality realism or non-realism locality? Recently, Simon
Groblacher et al \cite{zeil}, based on Leggett's work \cite{leg}, showed an
important conclusion that non-locality realism (i.e., give up locality but
keep in the realism) is still inconsistent with QM's predictions by both
theory and experiments. But this topic still need to be discussed further.

Although the two asummptions of LRT as shown above seem to be clear, they
are abstract when they are applied to real experimental situations. This is
the cause that one cannot judge easily whether or not there exists
non-locality realism or non-realism locality according to the assumptions.
In this work, we first discuss the detailed and operational desciption of
realism, and show that any system being in a pure state or a non-maximally
mixed state has property of non-realism. Then we present a strong and a weak
desciption of locality, and discuss the connections between non-locality and
entanglement and correlation. With the descriptions, one can easily find
theoretically and experimentally the inconsistence of locality and realism
with QM, and that both non-locality realism and non-realism locality are
should be given up in QM. We also present experimental schemes feasible
under current technologies to test the non-locality realsim.

Let's first consider the realism. EPR said that "We shall be satisfied with
the following criterion, which we regard as reasonable. \textit{If, without
in any way disturbing a system, we can predict with certainty (i.e., with
probability equal to unit) the value of a physical quantity, then there
exists an element of physical reality corresponding to this physical quantity%
}". They added that "Regarded not a necessary, but merely as a sufficient,
condition of reality, this criterion is in agreement with classical as well
as quantum-mechanical ideas of reality". EPR' criterion implies that the
values of physical reality elements can be predicted with certainty without
disturbing the systems. A observable should be a physical reality element.
So according to EPR's criterion and the description of realism in some other
papers \cite{chsh,cab1,zeil,0706}, the main idea of realism can be expressed
as follows: \textit{for a definite pure state }$\psi $\textit{\ of a system
one can predict with certainty the value of any observable }$A$\textit{\
without disturbing the system.}

If the state $\psi $\textit{\ }is not the eigenstate of the observable $A,$
QM claims that one can only predict the values $A_{1}$ or $A_{2}$ of
quantity $A$ with probability $p_{1}$ or $p_{2}$ (For simplicity,we suppose
A has only two eigenvalues)$,$ respectively, and the measurements will
disturb the system. This is not consistent with realism. To keep the
realism, hidden-variable model was suggested. HVM accept the probability
results of the meaurement of observable $A$ predicted by QM, but it thinks
that the realism is true and our no knowledge of the hidden-variable is the
cause of the probability results. According to the HVM, the property of the
system depends on not only the state $\psi $\textit{\ }but also a
hidden-variable $\lambda $ of which we now have no knowledge. It is to say
that the system may be in state $\psi (\lambda _{1})$ or $\psi (\lambda
_{2}) $ with probability $p_{1}$ or $p_{2},$ respectively. If the state of
the system is $\psi (\lambda _{1})$ $($or $\psi (\lambda _{2}))$ one can
predict with certainty the value $A_{1}$ (or $A_{2})$ of the observable $A$
without disturbing the system, i.e.,%
\begin{equation}
A\psi (\lambda _{1})=A_{1}\psi (\lambda _{1});\qquad A\psi (\lambda
_{2})=A_{2}\psi (\lambda _{2}).  \label{a}
\end{equation}%
According to HVM, the average value of the observable $A$ is $\overline{A}%
=A_{1}p_{1}+A_{2}p_{2}.$ According to QM, the average value $\left\langle
A\right\rangle =A_{1}p_{1}+A_{2}p_{2}.$ So if we only consider one
observable $A$, HVM and QM have the same prediction. It is to say that we
cannot judge whether HVM is true or not by the prediction of the average
value of one observable. Fortunately, we can judge HVM by investigating the
joint prediction of two non-commuting observables $A,B$ ($[A,B]=C\neq 0$) as
shown in the following Theorem 1.

Theorem 1: For any given state $\psi ,$\textit{\ }we can always find two
non-commuting observables $A$ and $B$ such that the average values of the
joint operators $AB$ and $BA$ can show the inconsistence between HVM and QM.

Proof: Without loss of generality, we introduce two hidden-variables $%
\lambda $ and $\lambda ^{\prime }$ to predict the values of observables $A$
and $B,$ respectively, where $B\psi (\lambda _{1}^{\prime })=B_{1}\psi
(\lambda _{1}^{\prime });B\psi (\lambda _{2}^{\prime })=B_{2}\psi (\lambda
_{2}^{\prime })$ similar to the observable $A$ in Eq. (\ref{a}). We suppose
that the system may be in the states $\psi (\lambda _{i},\lambda
_{j}^{\prime })$ with probability $p_{ij}$ $(i,j=1,2)$. According to HVM, we
can have that

\begin{eqnarray}
\overline{AB} &=&\sum_{i,j}p_{ij}A_{i}B_{j};  \notag \\
\overline{BA} &=&\sum_{i,j}p_{ij}B_{j}A_{i}=\overline{AB}  \label{b}
\end{eqnarray}%
Since $[A,B]=C\neq 0$, Eq. (\ref{b}) is not consistent with QM except for $%
\left\langle \psi \right\vert C\left\vert \psi \right\rangle =0.$ The left
proof of the theorem 1 is to prove that for a given state $\psi ,$ one can
always find two Hermian operators $A,B$ ($[A,B]=C\neq 0$) such that $%
\left\langle \psi \right\vert C\left\vert \psi \right\rangle \neq 0.$ To
this end, we can choose

\begin{eqnarray}
A &=&\left\vert \psi ^{\perp }\right\rangle \left\langle \psi \right\vert
+\left\vert \psi \right\rangle \left\langle \psi ^{\perp }\right\vert ;
\label{c} \\
B &=&-i(\left\vert \psi ^{\perp }\right\rangle \left\langle \psi \right\vert
-\left\vert \psi \right\rangle \left\langle \psi ^{\perp }\right\vert ); \\
C &=&2i(\left\vert \psi ^{\perp }\right\rangle \left\langle \psi ^{\perp
}\right\vert -\left\vert \psi \right\rangle \left\langle \psi \right\vert ),
\notag
\end{eqnarray}%
where $\left\vert \psi ^{\perp }\right\rangle $ is an arbitrary state
orthogonal to $\left\vert \psi \right\rangle $ of the system (The system can
be of higher dimensionality, but the operators $A,B$ and $C$ only belong to
the two-dimensional subspace of the whole space of the system). Thus, $%
\left\langle \psi \right\vert C\left\vert \psi \right\rangle =-2i\neq 0,$
which is not consistent with HVM as shown in Eq.(\ref{b}), and end the proof.

For a mixed state $\rho $, and two observables $A,B$ ($[A,B]=C\neq 0$),
according to HVM we still have $\overline{BA}=\overline{AB}.$ But by QM we
have

\begin{equation*}
\left\langle AB\right\rangle =Tr(\rho AB);\left\langle BA\right\rangle
=Tr(\rho BA).
\end{equation*}%
Suppose that the eigen-decomposition of density matrix $\rho $ is%
\begin{equation}
\rho =\sum_{i=1}^{n}p_{i}\left\vert \psi _{i}\right\rangle \left\langle \psi
_{i}\right\vert ,
\end{equation}%
where $\left\vert \psi _{i}\right\rangle s$ are orthogonal states and $%
\sum_{i=1}^{n}p_{i}=1,p_{i}\geqslant 0.$ If $\rho $ is not the maximally
mixed state $\rho _{\max }$ ($\rho _{\max }=\sum_{i=1}^{n}\frac{1}{n}%
\left\vert \psi _{i}\right\rangle \left\langle \psi _{i}\right\vert ),$ $%
\rho $ can be expressed as $\rho =\rho _{\max }+\sum_{i=1}^{n}\Delta
p_{i}\left\vert \psi _{i}\right\rangle \left\langle \psi _{i}\right\vert ,$
where $p_{i}=\frac{1}{n}+\Delta p_{i},\sum_{i=1}^{n}\Delta p_{i}=0.$ Without
loss of generality, let $\Delta p_{1}>0,\Delta p_{2}<0,$ then for $%
C=2i(\left\vert \psi _{1}\right\rangle \left\langle \psi _{1}\right\vert
-\left\vert \psi _{2}\right\rangle \left\langle \psi _{2}\right\vert ),$ $%
Tr(\rho C)=2i(\Delta p_{1}-\Delta p_{2})\neq 0.$ So if we choose $%
A=\left\vert \psi _{2}\right\rangle \left\langle \psi _{1}\right\vert
+\left\vert \psi _{1}\right\rangle \left\langle \psi _{2}\right\vert
;B=-i(\left\vert \psi _{1}\right\rangle \left\langle \psi _{2}\right\vert
-\left\vert \psi _{2}\right\rangle \left\langle \psi _{1}\right\vert ),$
then $Tr(\rho AB)-Tr(\rho BA)=Tr(\rho C)\neq 0$ for $\rho \neq \rho _{\max
}, $ which show the inconsistence between HVM and QM. So we can have the
following Theorem 2;

Theorem 2: For any non-maximally mixed state\textit{, }we can always find
two non-commuting observables $A$ and $B$ such that the average values of
the joint operators $AB$ and $BA$ can show the inconsistence between HVM and
QM.

If $\rho =\rho _{\max }\propto I,$ then for any observables $A$ and $B,$ $%
Tr(\rho AB)=Tr(\rho BA),$ which is consistent with QM. So we can conclude
that \textit{only the maximally mixed state always show the feature of
realism.}

Theorem 1 and 2 can be tested by currrent experimental technologies. For
example, let $\rho =p\left\vert 0\right\rangle \left\langle 0\right\vert
+(1-p)\left\vert 1\right\rangle \left\langle 1\right\vert ,$ where $%
\left\vert 0\right\rangle ,\left\vert 1\right\rangle $ are eigenstates of
the operator $\sigma _{z},$ i.e., spin-down and -up state or horizontal and
vertical polarization states of photon, respectively. If we let $A=\sigma
_{x};B=\sigma _{y},$ then we can test the Theorem 2 via measuring the
average values of the operators $\sigma _{x}\sigma _{y}$ and $\sigma
_{y}\sigma _{x}$ for the spin-1/2 particles or photon systems, where $%
(\sigma _{x},\sigma _{y},\sigma _{z})$ are Pauli operators.

We have shown the property of realism and its variance with QM in one
system. But almost all known Bell Theorems invlove two or more distant
systems rather than one system. How to describle the realism of the
composite system? Only if we regard the composite system as a whole system,
and the state and the observables $A,B$ in Eq. (\ref{c}) are those of the
whole system, then the description of realism and Theorems 1 and 2 can also
work. The key is to find a set of observables $A,B$ of the composite system
which can be measured easily under present teachnologies.

Now we provide a feasible scheme under our technologies to show the
realism's variance with QM in a bipartite system. The system under our
consideration is two spin-$\frac{1}{2}$ particles (or two polarization
photons) the state of which is $\psi =\cos \alpha \left\vert 00\right\rangle
+\sin \alpha \left\vert 11\right\rangle .$ let the observables $A=\sigma
_{x}^{(1)}\sigma _{x}^{(2)},B=(\overset{\rightharpoonup }{n}\bullet \overset{%
\rightharpoonup }{\sigma }^{(1)})(\overset{\rightharpoonup }{n}\bullet 
\overset{\rightharpoonup }{\sigma }^{(2)}),$ where $\overset{\rightharpoonup 
}{n}=(n_{x},n_{y},n_{z})$ is a three-dimensional unit vector. For $%
n_{x}=n_{y}=\frac{1}{\sqrt{2}},n_{z}=0,C=[A,B]=i(I\otimes \sigma
_{z}^{(2)}+\sigma _{z}^{(1)}\otimes I).$ According to HVM, the average
values $\overline{AB}=\overline{BA},$ but according to QM, $\left\langle
AB\right\rangle -\left\langle BA\right\rangle =\left\langle \psi \right\vert
C\left\vert \psi \right\rangle =i2\cos 2\alpha .$ Only we choose $\cos
2\alpha \neq 0,$ the difference of the expectations $\left\langle
AB\right\rangle $ and $\left\langle BA\right\rangle $ can show the realism's
variance with QM.

Let's now turn into locality. Locality claim that every measurement on A
system does not affect instantaneously the state of B system if A and B are
two distant systems. This means that measurements on the A system cannot
change the state of the B system in limit time. A strong description (or a
sufficient condition) of locality of a state $\rho _{AB}$ in AB system can
be expressed as: for \textit{any measurement,} describled by operators $%
\{M_{1},M_{2},\cdots .\dsum\limits_{i}M_{i}^{+}M_{i}=I\},$ on A system,
whatever the outcome $i$ is, the corresponding state $\rho _{iB}$ of the B
system is always $\rho _{B},$ where $\rho _{iB}=Tr_{A}(M_{i}\rho
_{AB}M_{i}^{+})$ and $\rho _{B}=Tr_{A}(\rho _{AB}).$ While a weak
description (or a necessary condition) can be as: there exists \textit{a
measurement} on A system such that for each outcome $i$ the state of the B
system is always $\rho _{B}.$

Theorem 3: For a pure state $\psi _{AB}$, the system satisfies both the
strong and the weak locality if and only if its state is separable.

Proof: If the state is pure and separable, the system satisfies obviously
both the strong and the weak locality. If the state $\psi _{AB}$ is
entangled, then for any measuement denoted by operators $\{M_{1},M_{2},%
\cdots .\dsum\limits_{i}M_{i}^{+}M_{i}=I\}$ on the A system there always
exists a outcome $i$ such that the B system will collapse correspondlly into
a state not being $\rho _{B}$ (except that $M_{i}$ is proprotional to a unit
operator, which means the measurement is trivial). End of the proof.

For mixed states, some states only satisfy the weak description, but do not
satisfy the strong description of locality. For example, for the classical
correlated state $\rho _{AB}=\frac{1}{2}\left\vert 00\right\rangle
\left\langle 00\right\vert +\frac{1}{2}\left\vert 11\right\rangle
\left\langle 11\right\vert ,$ if the measurement operators on the A system
are $M_{1}=\left\vert 0\right\rangle \left\langle 0\right\vert
,M_{2}=\left\vert 1\right\rangle \left\langle 1\right\vert ,$ the state of
the B system will be collapsed into $\rho _{1B}=\left\vert 0\right\rangle
\left\langle 0\right\vert ,\rho _{2B}=\left\vert 1\right\rangle \left\langle
1\right\vert $ corresponding to the outcome 1 and 2, respectively, which are
not $\rho _{B}=\frac{1}{2}\left\vert 0\right\rangle \left\langle
0\right\vert +\frac{1}{2}\left\vert 1\right\rangle \left\langle 1\right\vert 
$. But if the measurement operators on the A system are $M_{1}=\frac{1}{2}%
\left\vert 0+1\right\rangle \left\langle 0+1\right\vert ,M_{2}=\frac{1}{2}%
\left\vert 0-1\right\rangle \left\langle 0-1\right\vert ,$ both $\rho _{1B}$
and $\rho _{2B}$ are the same states as $\rho _{B}.$ So the state $\rho
_{AB} $ only satisfies the weak description of locality.

Theorem 4 (Strong locality Theorem): The state $\rho _{AB}$ of the AB system
satisfies the strong description of locality if and only if the $\rho _{AB}$
can be expressed as the form of $\rho _{AB}=\rho _{A}\otimes \rho _{B}$.

Proof: If $\rho _{AB}=\rho _{A}\otimes \rho _{B},$ it is obvious that the
state $\rho _{AB}$ satisfies the strong description of locality. If $\rho
_{AB}\neq \rho _{A}\otimes \rho _{B},$ there exists correlation between the
system A and B. So one can acquire some information of the B system via a
measurment on the A system. This means that there is at least one measurment
expressed by the operators $\{M_{1},M_{2},\cdots
.\dsum\limits_{i}M_{i}^{+}M_{i}=I\}$ on the A system such that at least one
state $\rho _{iB}$ of the B system corresponding to measurement outcome $i$
is not $\rho _{B}$ (The acquired information of the B system via the
measurement on the A system is $S(\rho _{B})-S(\rho _{iB}),$ where $S(.)$ is
von Neumann entropy. If $S(\rho _{B})-S(\rho _{iB})\neq 0,$ then $\rho
_{iB}\neq \rho _{B}),$ and so the state $\rho _{AB}$ does not satisfy the
strong description of locality. End of the proof.

Theorem 4 shows any correlated states do not satisfy the strong description
of locality.

For the weak description, we have following Theorem:

Theorem 5: Suppose A is a two-dimension system, and B is of arbitary
dimension. For a state $\rho _{AB}$ of the AB system, if there exists a set
of projective measurement expressed by operators $\{P_{Ai},i=1,2,\dsum%
\limits_{i=1}^{2}P_{Ai}=I\}$ on A system such that the corresponding state
of the B system is always $\rho _{B}$ for each outcome $i$, then the state $%
\rho _{AB}$ is separable.

Proof: Without loss of generality, we can imagine that AB is a subsystem of
a big composite system ABC, the state of the composite system is%
\begin{equation}
\psi _{ABC}=\dsum\limits_{k=1}^{n}\sqrt{p_{k}}\left\vert \psi
_{AB}^{k}\right\rangle \left\vert k\right\rangle _{C},
\end{equation}%
where $\left\vert k\right\rangle _{C}s$ is a set of orthogonal states of the
C system, $\{p_{k},\psi _{AB}^{k}\}$ is a set of eigenstate-decomposition of
the density matrix $\rho _{AB}.$ If there exists a set of projective
measurement operators $\{P_{Ai}=\left\vert \phi _{i}\right\rangle
\left\langle \phi _{i}\right\vert ,i=1,2,\dsum\limits_{i=1}^{2}P_{Ai}=I\}$
on A system such that the state of B system always is $\rho _{B}$ for each
outcome $i$, where $\{\left\vert \phi _{1}\right\rangle ,\left\vert \phi
_{2}\right\rangle \}$ is a set of bases of the A system, then $\psi _{ABC}$
can be expressed as 
\begin{eqnarray}
\psi _{ABC} &=&\dsum\limits_{i=1}^{2}P_{Ai}\psi _{ABC}=  \notag \\
&&\sqrt{\lambda _{1}}\left\vert \phi _{1}\right\rangle _{A}(\left\vert \eta
_{11}\right\rangle _{B}\left\vert 1\right\rangle _{C}+\ldots +\left\vert
\eta _{1n}\right\rangle _{B}\left\vert n\right\rangle _{C})+  \notag \\
&&\sqrt{\lambda _{2}}\left\vert \phi _{2}\right\rangle _{A}(\left\vert \eta
_{21}\right\rangle _{B}\left\vert 1\right\rangle _{C}+\ldots +\left\vert
\eta _{2n}\right\rangle _{B}\left\vert n\right\rangle _{C}),
\end{eqnarray}%
and for each outcome $i$ the states of B system $\rho
_{Bi}=\dsum\limits_{k=1}^{n}\left\vert \eta _{ik}\right\rangle \left\langle
\eta _{ik}\right\vert ,i=1,2,$ are equal to $\rho _{B},$ where $\left\vert
\eta _{ik}\right\rangle _{B},i=1,2;k=1,\cdots ,n,$ are unnormalized vectors
of the B system. Two sets of states $\{\left\vert \eta _{i1}\right\rangle
,\cdots ,\left\vert \eta _{in}\right\rangle \},i=1,2,$ are two sets of the
pure-state-decompositions of the density matrix $\rho _{B},$ so they can be
connected by a unitary matrix \cite{wott} as 
\begin{equation}
\left[ U_{0}\right] \left[ 
\begin{array}{c}
\left\vert \eta _{11}\right\rangle  \\ 
\vdots  \\ 
\left\vert \eta _{1n}\right\rangle 
\end{array}%
\right] =\left[ 
\begin{array}{c}
\left\vert \eta _{21}\right\rangle  \\ 
\vdots  \\ 
\left\vert \eta _{2n}\right\rangle 
\end{array}%
\right] ,  \label{a1}
\end{equation}%
where $U_{0}$ is a unitary matrix. By linear algebra, any unitary matrix $%
U_{0}$ can be diagolized by a unitary matrix $U,$ i.e.,%
\begin{equation}
UU_{0}U^{-1}=\Lambda ,  \label{a2}
\end{equation}%
where $\Lambda $ is a diagonal matrix and its diagonal elements $\Lambda
_{kk}(k=1,...,n)$ are of norm 1 (i.e., $\left\vert \Lambda _{kk}\right\vert
=1).$ On the other hand, considering another set of bases of the C system $%
\{\left\vert 1^{\prime }\right\rangle _{C},\ldots ,\left\vert n^{\prime
}\right\rangle _{C}\}$ such that the transformtion matrix from the bases $%
\{\left\vert 1\right\rangle _{C},\ldots ,\left\vert n\right\rangle _{C}\}$
to bases $\{\left\vert 1^{\prime }\right\rangle _{C},\ldots ,\left\vert
n^{\prime }\right\rangle _{C}\}$ is $U,$ we can re-express the state $\psi
_{ABC}$ as%
\begin{eqnarray}
&&\psi _{ABC}  \notag \\
&=&\sqrt{\lambda _{1}}\left\vert \phi _{1}\right\rangle _{A}(\left\vert \eta
_{11}^{\prime }\right\rangle _{B}\left\vert 1^{\prime }\right\rangle
_{C}+\ldots +\left\vert \eta _{1n}^{\prime }\right\rangle _{B}\left\vert
n^{\prime }\right\rangle _{C})+  \notag \\
&&\sqrt{\lambda _{2}}\left\vert \phi _{2}\right\rangle _{A}(\left\vert \eta
_{21}^{\prime }\right\rangle _{B}\left\vert 1^{\prime }\right\rangle
_{C}+\ldots +\left\vert \eta _{2n}^{\prime }\right\rangle _{B}\left\vert
n^{\prime }\right\rangle _{C}),  \label{a3}
\end{eqnarray}%
where%
\begin{equation}
\left[ U\right] \left[ 
\begin{array}{c}
\left\vert \eta _{i1}\right\rangle  \\ 
\vdots  \\ 
\left\vert \eta _{in}\right\rangle 
\end{array}%
\right] =\left[ 
\begin{array}{c}
\left\vert \eta _{i1}^{\prime }\right\rangle  \\ 
\vdots  \\ 
\left\vert \eta _{in}^{\prime }\right\rangle 
\end{array}%
\right] ,i=1,2.  \label{a4}
\end{equation}%
Combinating Eqs. (\ref{a1}, \ref{a2} \ref{a4}), we have that 
\begin{equation*}
\left[ 
\begin{array}{c}
\left\vert \eta _{21}^{\prime }\right\rangle  \\ 
\vdots  \\ 
\left\vert \eta _{2n}^{\prime }\right\rangle 
\end{array}%
\right] =\left[ \Lambda \right] \left[ 
\begin{array}{c}
\left\vert \eta _{11}^{\prime }\right\rangle  \\ 
\vdots  \\ 
\left\vert \eta _{1n}^{\prime }\right\rangle 
\end{array}%
\right] .
\end{equation*}%
So Eq. (\ref{a3}) can be expressed as%
\begin{eqnarray}
&&\psi _{ABC}  \notag \\
&=&(\sqrt{\lambda _{1}}\left\vert \phi _{1}\right\rangle _{A}+\Lambda _{11}%
\sqrt{\lambda _{2}}\left\vert \phi _{2}\right\rangle _{A})\left\vert \eta
_{11}^{\prime }\right\rangle _{B}\left\vert 1^{\prime }\right\rangle _{C}+ 
\notag \\
&&\ldots +(\sqrt{\lambda _{1}}\left\vert \phi _{1}\right\rangle _{A}+\Lambda
_{nn}\sqrt{\lambda _{2}}\left\vert \phi _{2}\right\rangle _{A})\left\vert
\eta _{1n}^{\prime }\right\rangle _{B}\left\vert n^{\prime }\right\rangle
_{C}).  \label{a5}
\end{eqnarray}%
Obviously, from Eq. (\ref{a5}) we can see that the state $\rho _{AB}$ is
separable, and end the proof.

Theorems 3, 4 and 5 imply that the entangled states and correlated states
might show the variance between locality and the QM's prediction.

In summary, we present the detailed and applicable desciptions of realism
and locality in the real experimental situations. With the descriptions, one
can easily find theoretically and experimentally the inconsistence of
locality and realism with QM. We also present experimental schemes feasible
under current technologies to test the non-locality realsim. However, there
are some open questions need to be discussed further. For example, what is
the difference between the non-realism and the non-locality in a single
particle as shown in Refs \cite{tan,hardy1,ved}?

This work is supported by the National Natural Science foundation of China
(No. Grants 10404039 and 05-06/01), a Foundation for the Author of National
Excellent Doctoral Dissertation of China No. 200524 and NCET of China.

\end{document}